\documentclass[twocolumn,amssymb,fleqn]{revtex4} 
\usepackage{epsfig,amssymb,amsmath,graphicx,subfigure,hyperref}  
\usepackage{color}

\usepackage{multirow} 
\usepackage{tabularx}

\begin{document}

\title{Fast dynamics and emergent topological defects in long-range \\ interacting
particle systems} \author{Zhenwei Yao} \email{zyao@sjtu.edu.cn}
\affiliation{School of Physics and Astronomy, and Institute of Natural Sciences,
Shanghai Jiao Tong University, Shanghai 200240, China}

\begin{abstract}
Long-range interacting systems exhibit unusual physical properties not shared
  by systems with short-range interactions. Understanding the dynamical and
  statistical effects of long-range interactions yields insights into a host of
  physical systems in nature and industry. In this work, we investigate the
  classical microscopic dynamics of screened Coulomb interacting particles
  confined in the disk, and reveal the featured dynamics and emergent
  statistical regularities created by the long-range interaction. We highlight
  the long-range interaction driven fast single-particle and collective
  dynamics, and the emergent topological defect structure. This work suggests
  the rich physics arising from the interplay of long-range interaction,
  topology and dynamics. 
\end{abstract}

\maketitle

\section{Introduction}

The classical mechanical model of interacting particles is of great historical
and scientific
significance~\cite{cercignani1998ludwig,boltzmann1964lectures,Ma1985}. Based on
the short-range interacting particle model, Ludwig Boltzmann completed the
statistical interpretation of thermodynamics, and derived the famous $H$-theorem
that has inspired profound discussions on the foundation of statistical
mechanics~\cite{boltzmann1964lectures,ehrenfest2002conceptual,frenkel2016interview}.
Generalizing the physical interaction to the long-range regime invalidates the
basic concepts of additivity and extensivity, which constitute the foundation of
thermodynamics~\cite{campa2014physics,levin2014nonequilibrium,pakter2017entropy}.
Long-range interactions are also widely seen in a variety of interdisciplinary
systems at length scales covering multiple orders of
magnitude~\cite{lighthill1976flagellar,chattopadhyay2009effect,tabi2010long,dallaston2018discrete,yao2019command}.  Especially, the long-range nature of the
electrostatic interaction is crucial for the self-assembly of exceedingly rich
soft matters in electrolyte
solutions~\cite{Holm2001,Levin2002,Walker2011,toor2016self,gao2019electrostatic}.
To deal with the notoriously challenging long-range interacting many-body
systems, the approach based on numerical integration of the equations of motion
at high precision has proven to be a powerful tool to reveal the fundamental
microscopic dynamics not accessible by mean-field
theories~\cite{rapaport2004art,campa2014physics}.  Elucidating the dynamical
effects of long-range interactions yields insights into various nonequilibrium
processes ranging from
hydrodynamics~\cite{lighthill1976flagellar,chattopadhyay2009effect,dallaston2018discrete}
to electrostatic
self-assembly~\cite{grzybowski2003electrostatic,Walker2011,vernizzi2011coulomb,lindgren2018electrostatic}.

The goal of this work is to investigate the classical dynamics of screened
Coulomb interacting particles confined in the disk. This work represents a
generalization of the previous studies on the equilibrium packing of charged
point particles in the disk from the static to the dynamical
regime~\cite{Mughal2007,yao2013topological,soni2018emergent}. This theoretical
model can be realized by colloidal experiments, where the screening length is
tunable by the salt concentration~\cite{debye1923theory,Dobrynin2005,Holm2001,soni2018emergent}. Previous studies of the static disk model
show that the inhomogeneity in density created by the long-range repulsion could
induce Gaussian curvature and excite topological
defects~\cite{nelson2002defects,Mughal2007,yao2013topological,soni2018emergent}.
As a fundamental topological defect, a disclination in a
triangular lattice refers to a vertex whose coordination
number $z$ deviates from six. The disk model provides the opportunity
to clarify a host of questions with broad implications, such as: What is the
distinction of short- and long-range interactions in commanding single-particle
and collective dynamics? Will the defect structure revealed in the static system
still persist in the dynamical regime, and if yes, in which form?

To address these fundamental questions, we resort to the adaptive Verlet method
to construct long-time, energy-conserved particle
trajectories~\cite{rapaport2004art}, and reveal the featured energy transfer
mode and fast single-particle and collective dynamics under the long-range
interaction. By analyzing the convoluted collective dynamics from the unique
perspective of topological defects, we identify the emergent statistical
regularity in the distribution of topological charges, and uncover the
fundamentally different defect structures in short- and long-range interacting
systems. The discovery of the emergent dynamical and statistical regularities
in this work may have implications in characterizing the intriguing physical effects of
long-range interactions.

\begin{figure*}[ht]  
\centering 
\includegraphics[width=6.8in, bb=38 102 680 518]{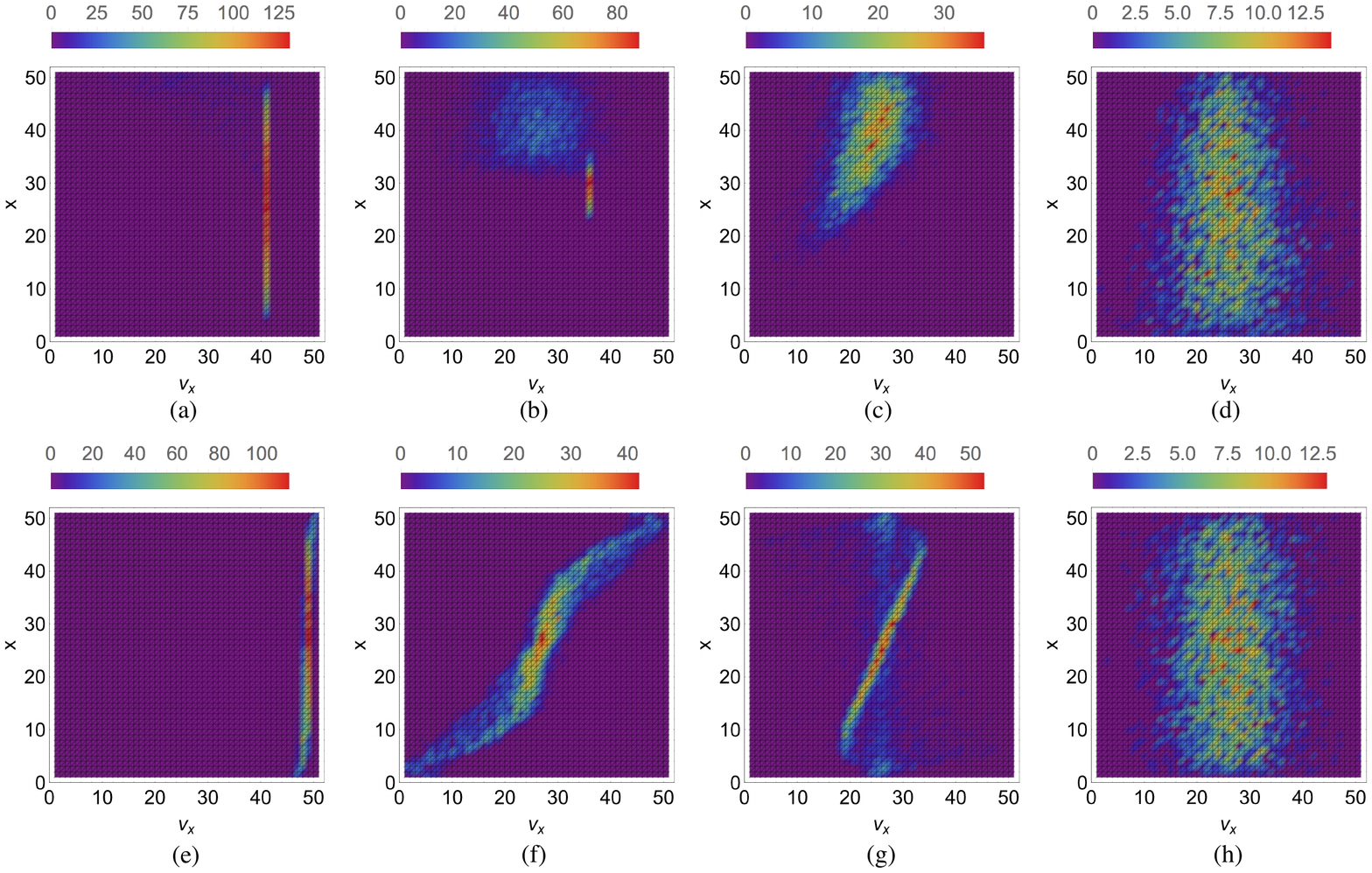}
\caption{Short- and long-range interacting particle systems exhibit distinct
  scenarios of dynamical evolution as represented in the space of $\{x,
  v_x\}$. (a)-(d) $\lambda/a=0.05$. $t/\tau_0=\{0.15, 0.90, 1.35, 5.85\}$.
  (e)-(h) $\lambda/a=10$. $t/\tau_0=\{0.0015, 0.009, 0.036, 6.0\}$. The 
  space is divided into $50\times 50$ cells. The colored legends indicate the
  number of particles in $\delta x \delta v_x$. $x \in [-x_m, x_m]$ and $v_x \in
  [-v_m, v_m]$. (a)-(d) $x_m= \{1.0, 1.0, 1.0, 1.0\}$. $v_m=\{ 1.73, 2.5, 2.7,
  3.6\}$.  (e)-(h) $x_m= \{1.0, 1.12, 1.13, 1.15\}$. $v_m= \{1.12, 27.1, 44.7,
  58.2 \}$. $k_0=10^5$. $N=5000$. $V_0=1$.
   }
\label{fig_phase}
\end{figure*}

\begin{figure}[th]  
\centering 
\includegraphics[width=3.4in]{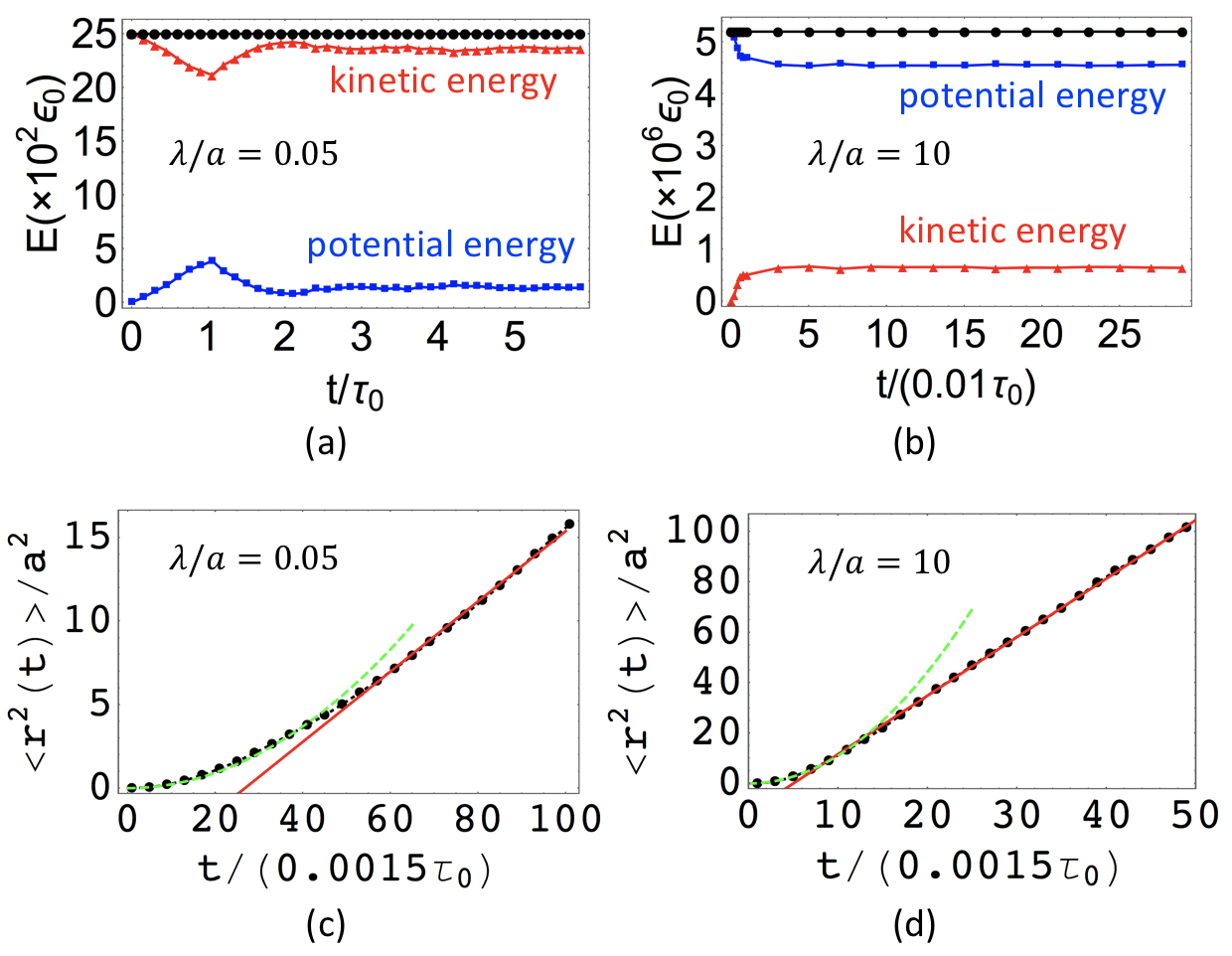}
  \caption{Fast dynamics under the long-range interaction as demonstrated in the
  energy transfer process and the single-particle motion. (a) and (b) Temporal
  variation of the system energy for typical short- and long-range interacting
  systems.  (c) and (d) Plots of the mean squared displacement of a single
  particle near the center of the disk. The quadratic (dashed green) and linear
  (solid red) fitting curves indicate the ballistic and diffusive motions. 
  $N=5000$. $k_0=10^5$. $V_0=1$. }
\label{fig_energy}
\end{figure}

\begin{figure*}[thb]  
\centering 
\includegraphics[width=7in]{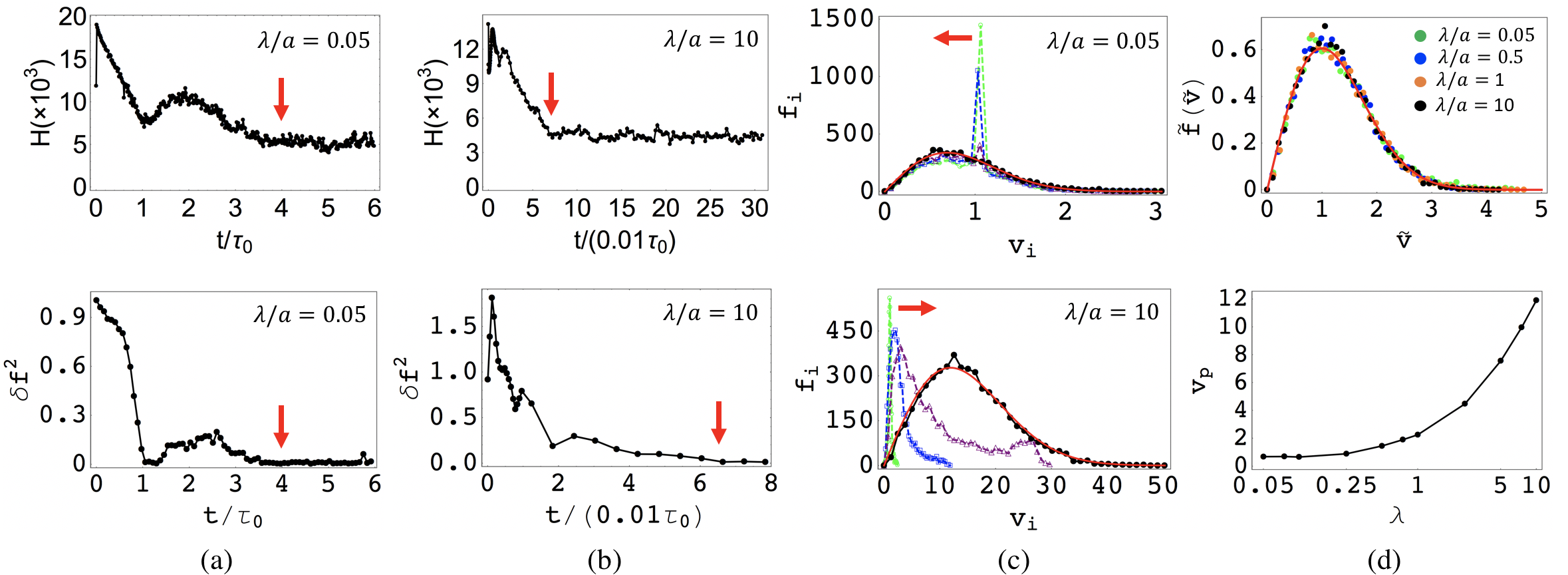}
  \caption{Fast relaxation of the particle speed under the long-range
  interaction. (a) $\lambda/a=0.05$. (b) $\lambda/a=10$. The relaxation process
  is characterized by both the $H$ function and the speed distribution function.
  $\delta f^2$ is the deviation of the speed distribution function $f(v)$ from
  the Maxwell-Boltzmann distribution (see the main text for more information).
  The red arrows in (a) and (b) indicate the common relaxation time as read from
  both the $H$- and $\delta f^2$-curves. (c) Typical instantaneous
  distribution curves in the relaxation process towards the equilibrium
  distribution (solid red curves). Upper panel: $t/\tau_0= 0.77$ (green),
  0.86 (blue), 1.0 (purple), and 5.85 (black). Lower panel: $t/\tau_0= 0.00012$
  (green), 0.001 (blue), 0.004 (purple), 6.0 (black). The red arrows indicate
  the opposite movement of the peak. (d) Uniform collapse of the reduced speed
  distributions at varying $\lambda$ on the standard normalized
  Maxwell-Boltzmann distribution. The plot of the most probable speed $v_p$
  against $\lambda$ is presented in the lower panel. $N=5000$. $k_0=10^5$. $V_0=1$. }
\label{fig_speed}
\end{figure*}

\begin{figure}[thb]  
\centering 
\includegraphics[width=3.4in]{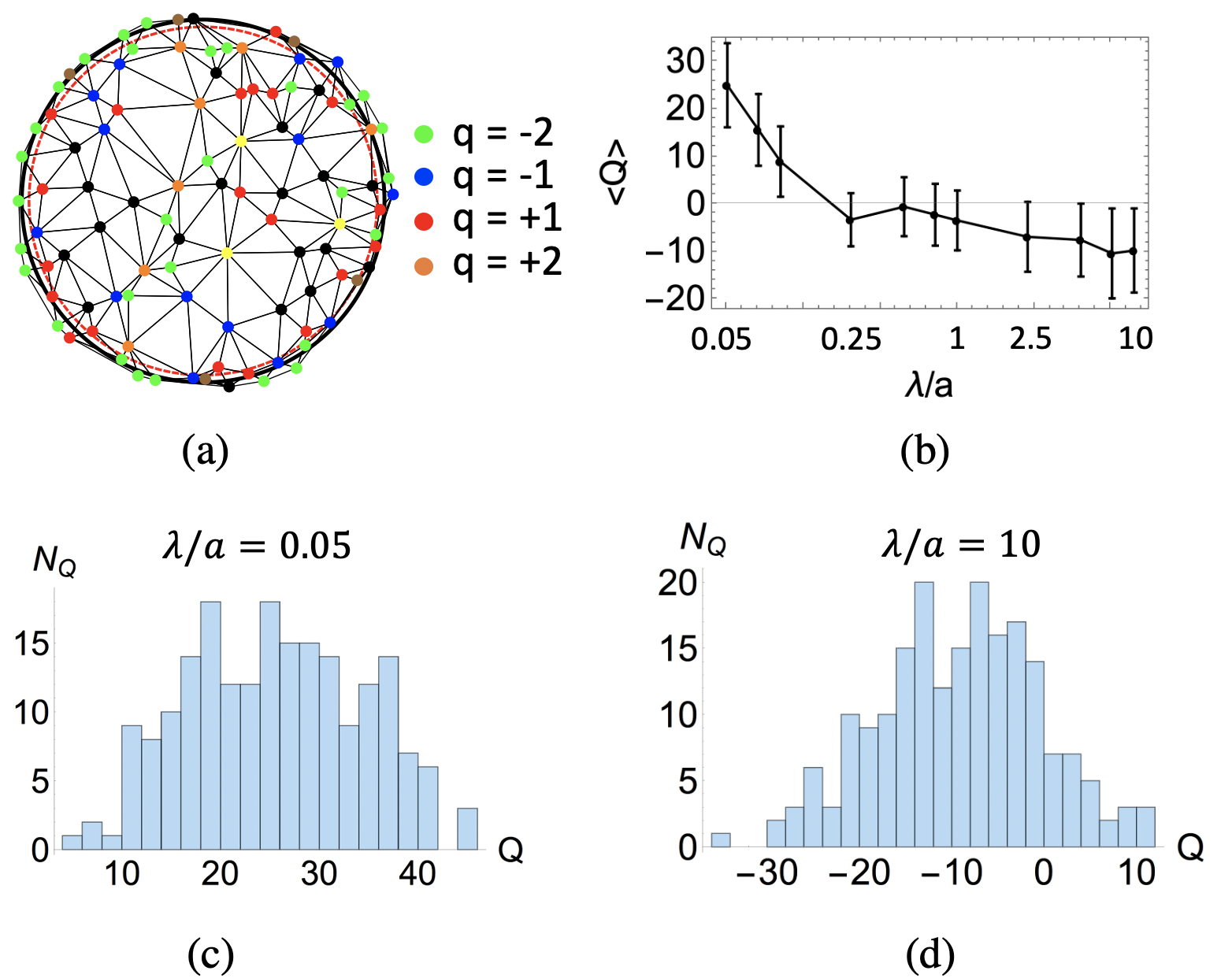}
  \caption{Short- and Long-range interacting systems exhibit fundamentally
  different topological defect structures. (a) Classification of the particles
  by the topological charge. (b) Plot
  of $\langle Q \rangle$ vs $\lambda$. $\langle Q \rangle$ is the mean total topological charge
  within the circular region indicated by the dashed red circle in (a). Notice
  that $\langle Q \rangle$ becomes negative at large $\lambda$. (c) and (d) show
  the distributions of $Q$ in statistically independent equilibrium particle
  configurations for $\lambda/a=0.05$ and $\lambda/a=10$, respectively.
  $N=5000$. $k_0=10^5$. $V_0=1$.
    }
\label{fig_defect}
\end{figure}

\section{Model and method}

The model consists of a collection of $N$ identical point particles of mass $m_0$ confined in a disk of
radius $r_0$ interacting by the screened Coulomb potential. The Hamiltonian of
the system is
\begin{eqnarray}
  H=\sum_{i=1}^{N} \left(\frac{\vec{p}_i^2}{2m}+V_{\textrm {wall}}(|\vec{x}_i|)\right) + \sum_{i\neq
  j}V(|\vec{x}_i-\vec{x}_j|).
  \label{H}
\end{eqnarray}
The confining potential
\begin{eqnarray}
  V_{\textrm {wall}}(|\vec{x}_i|) = \frac{1}{2}k_0
(|\vec{x}_i|-r_0)^2 \chi(|\vec{x}_i|-r_0), 
\end{eqnarray}
  where $\chi(x)=1$ if $x>0$, and
$\chi(x)=0$
otherwise. The screened Coulomb potential 
\begin{eqnarray}
  V(r)= V_0  \frac{1}{r} e^{-\frac{r}{\lambda}}. \label{debye}
  \end{eqnarray}
The screened Coulomb potential allows us to investigate the effects of the
long-range interaction by comparison with that of the short-range interaction.
Note that the charge neutrality condition requires the existence of opposite
charges, which are treated as continuum background charges in our model.
The model system of charged particles with the effective interaction in the form
of Eq.(\ref{debye}) could be experimentally realized by charged polymethyl
methacrylate (PMMA) particles in electrolyte
solutions~\cite{soni2018emergent,chen2020morphology}. 
No cut-off length is introduced in the calculation of the long-range interaction
force. We distinguish the short- and long-range interactions by the ratio
$\lambda/a$ by physical consideration,
where $a$ is the mean distance between nearest particles. Under the assumption
that the particles are arranged by triangular lattice, it is estimated that
$a/r_0=\sqrt{2\pi/\sqrt{3}N}$.

The dynamical evolution of the system is governed by the Hamiltonian in
Eq.(\ref{H}). We construct long-time, energy-conserved particle trajectories by
the adaptive Verlet method, where the time step is dynamically varying for
striking a balance of the energy conservation and the computational efficiency.
To highlight the effect of the interaction range on the relaxation process,
the particles are specified with a uniform velocity $\vec{v}(t=0)=v_0\hat{x}$ in
the initial state. The initial positions of the particles are randomly
distributed by the standard procedure of random disk
packing~\cite{lubachevsky1990geometric}. We analyze the relaxation process from
both perspectives of the $H$-function and the speed distribution function. In this
work, the units of length, mass, time, and energy
are $r_0$, $m_0$, $\tau_0$, and $\epsilon_0$, respectively.  $\epsilon_0 = m_0
(r_0/\tau_0)^2$. $\tau_0 = r_0/v_0$.  Typically, $h=10^{-7}$ and $k_0=10^5$ to
ensure that the total energy is well conserved and the particles are
geometrically confined in the disk in our simulations. More simulation details
are presented in Appendix A. The effects of the stiffness $k_0$ of the confining
potential on both the relaxation rate and the particle density distribution are
discussed in Appendices B and C.

\section{Results and discussion}

In Fig.~\ref{fig_phase}, we show typical instantaneous states in the dynamical evolution of
the system in the space spanned by the parameters $x$ and $v_x$ for
$\lambda/a=0.05$ [Figs.~\ref{fig_phase}(a)-(d)] and $\lambda/a=10$
[Figs.~\ref{fig_phase}(e)-(h)]. The dynamical evolutions in the complementary
space of $\{y, v_y\}$ are presented in Appendix B. $\{x, y, v_x, v_y\}$
constitute a complete single-particle phase space ($\mu$-space), where each
point represents one particle with specific position and velocity. The density
of the points in the $\mu$-space is indicated by color. We see that the short-
and long-range interacting systems exhibit distinct dynamical behaviors.  The
system of $\lambda/a=10$ reaches the equilibrium state much faster than that of
$\lambda/a=0.05$. We also notice that, for the case of $\lambda/a=0.05$, the
initial uniform motion of the particles along x-axis leads to the shrinking of
the occupied belt-like region, as shown in Fig.~\ref{fig_phase}(b). In
contrast, the disk is always fully occupied for the case of $\lambda/a=10$.

The total mechanical energy of the system is well conserved in the dynamical
evolution of the system, as shown in Figs.~\ref{fig_energy}(a) and
\ref{fig_energy}(b). We see that the short- and long-range interacting systems
exhibit distinct energy transfer mode between the kinetic and potential
energies.  Figure~\ref{fig_energy}(b) shows the simultaneous reduction of the
potential energy and the increase of the kinetic energy, indicating the
conversion of the potential energy to the kinetic energy. In contrast, we find
that the situation is opposite for the case of $\lambda/a=0.05$ in
Fig.~\ref{fig_energy}(a). This featured energy transfer scenario is uniformly
observed by changing the value of $V_0$ in the screened Coulomb potential from
$V_0=0.1$ to $V_0 = 5$ (see Fig.~\ref{fig_energy_V0} in Appendix B). This
observation implies that, for long- and short-range interacting systems with
identical initial state, the former system may possess a higher temperature in
the final equilibrium state, which will be discussed later. More information
about the dependence of the potential and kinetic energies on $\lambda$ is
presented in Appendix A.  Figures~\ref{fig_energy}(a) and \ref{fig_energy}(b)
also show that the kinetic and potential energy curves ultimately become flat,
which can be attributed to concurrent inverse collision processes in the
many-particle system.

The fast relaxation of the long-range interacting system, as observed in
Fig.~\ref{fig_phase}, is also reflected in the variation of the energy curves in
Figs.~\ref{fig_energy}(a) and \ref{fig_energy}(b); note the different time
scales in the abscissa axes. A question naturally arises: Do individual
particles move faster under the long-range interaction? To address this
question, we analyze the dynamics of a randomly picked single particle near the
center of the disk to avoid any boundary effect. By averaging over 100
statistically independent particle trajectories, we plot the mean squared
displacement $\langle r^2(t) \rangle$ in Figs.~\ref{fig_energy}(c) and
\ref{fig_energy}(d) for both cases of $\lambda/a=0.05$ and $\lambda/a=10$. We
uniformly observe the shift of the initial ballistic motion (the
dashed green quadratic fitting curves) to the diffusive motion (the solid red
linear fitting lines).  Remarkably, the diffusion coefficient for $\lambda/a=10$
is about 11 times of that for $\lambda/a=0.05$. The speed of the initial
ballistic motion for the former system is about 48 times faster than the latter
one. To conclude, the single particle dynamics is significantly faster under
the long-range interaction.

We proceed to discuss the influence of the range of interaction on the
collective dynamics. Specifically, we focus on the relaxation of the particle
speed in the short- and long-range interacting systems. According to 
classical statistical physics, if the interaction potential is a function of particle coordinates
only, the distribution of the particle speed uniformly
conforms to the Maxwell-Boltzmann distribution regardless of the range of
interaction~\cite{Ma1985}. However, the range of interaction may affect the
kinetic pathway of the relaxation process. To clarify this question, we
quantitatively characterize the relaxation process by both the $H$ function
and the quantity $\delta f^2$. As a measure of the deviation of the instantaneous
speed distribution $f(v)$ from the
Maxwell-Boltzmann distribution $f_{\textrm{eq.}}$, $\delta f^2$ is defined as
\begin{eqnarray}
  \delta f^2 = \int_0^{\infty }(f(v)-f_{\textrm{eq.}}(v))^2 dv.
  \end{eqnarray}
The $H$ function measures the relative probability of an out-of-equilibrium
state, and its discrete expression is~\cite{boltzmann1964lectures}:
\begin{eqnarray} 
  H=\sum_i n_i \log n_i,
\end{eqnarray} 
where $n_i$ is the number of particles in the $i-$th cell of the $\mu$-space spanned
by $\{x, y, v_x, v_y\}$.

Figures~\ref{fig_speed}(a) and \ref{fig_speed}(b) show the $H$- and $\delta
f^2$-curves for the cases of $\lambda/a=0.05$ and $\lambda/a=10$, respectively.
From the $H$-curves, we see that both systems tend to evolve along the direction of
reducing $H$~\cite{Ma1985}. The $H$-curves are subject to persistent
fluctuations in equilibrium. One could read the relaxation time from both the
$H$- and $\delta f^2$-curves. These curves become stable after the
characteristic sites as indicated by the red arrows in Figs.~\ref{fig_speed}(a)
and \ref{fig_speed}(b). And these sites give the value for the relaxation time.
The relaxation times as read from both the $H$- and $\delta f^2$-curves are
identical, which implies the consistency of both quantities of $H$ and
$\delta f^2$ to characterize the relaxation process. Furthermore, comparison of
Figs.~\ref{fig_speed}(a) and \ref{fig_speed}(b) shows the significantly faster
relaxation rate under the long-range interaction by two orders of magnitude.  We
also observe much faster relaxation of the orientation of particle velocity in
larger-$\lambda$ systems (see Appendix B for more information).

Typical instantaneous speed distributions are presented in
Fig.~\ref{fig_speed}(c). We see that the short- and long-range interacting
systems exhibit distinct kinetic pathways in the relaxation process. The peak in
the distribution profile moves along opposite directions in these two kinds of
systems as indicated by the arrows. This phenomenon can be understood from
Fig.~\ref{fig_energy}. Figure~\ref{fig_energy}(b) shows that the potential
energy is converted into kinetic energy for the case of $\lambda/a=10$. The
increase of the kinetic energy leads to the rightward movement of the peak in
the lower panel for $\lambda/a=10$ in Fig.~\ref{fig_speed}(c). This process is
reversed for the case of $\lambda/a=0.05$. Furthermore, in the relaxation of the
long-range interacting system, we notice the appearance of a new peak in the
tail of the distribution profile [see the purple curve in the lower panel in
Fig.~\ref{fig_speed}(c)]. This peak originates from the accumulation of particle
population in the high speed regime; the associated extra kinetic energy is
provided by the release of the stored potential energy in the long-range
repulsive system [see Fig.~\ref{fig_energy}(b)].

The resulting equilibrium speed distributions indicated by the solid red
curves in Fig.~\ref{fig_speed}(c) can be well fitted by the
two-dimensional Maxwell-Boltzmann distribution:
\begin{eqnarray}
  f(v)\delta v = N\frac{v}{v_p^2}\exp\left(-\frac{v^2}{2v_p^2}\right) \delta v,
  \label{eq_fv}
\end{eqnarray}
where the most probable speed $v_p = \sqrt{k_BT/m}$. By rescaling the speed by
$v_p$, all the reduced speed distributions at varying $\lambda$ uniformly
collapse on the standard normalized Maxwell-Boltzmann distribution, as shown in
Fig.~\ref{fig_speed}(d). The dependence of $v_p$ on $\lambda$ is shown in the
lower panel in Fig.~\ref{fig_speed}(d). Since $T \propto v_p^2$, we conclude that under the
same initial condition the temperature of a larger-$\lambda$ system in
equilibrium is indeed significantly higher.

Now, we analyze the collective dynamics from the perspective of the underlying
topological defect structure. For a two-dimensional particle array, one can
resort to the standard Delaunay triangulation procedure to identify the
particles whose coordination number $z$ is deviated from
six~\cite{nelson2002defects}. These particles are known as disclinations,
carrying topological charge $q=6-z$.  Elasticity theory shows a remarkable
analogy of disclinations and electric charges; oppositely charged disclinations
attract and like-signs repel.  The concept of topological charge has proven
crucial for understanding 2D crystal
melting~\cite{Kosterlitz1973,halperin1978theory}, healing of crystalline
order~\cite{irvine2012fractionalization,yao2014polydispersity}, packing of
twisted filament bundles and virus~\cite{kamien1995iterated,Grason2010,lidmar2003virus}, and non-equilibrium
dynamics~\cite{marchetti2013hydrodynamics,keber2014topology}.

In Fig.~\ref{fig_defect}(a), we demonstrate the Delaunay triangulation of an
instantaneous particle configuration. Different types of
disclinations are indicated with different colors. We focus on the total
topological charge $Q$ within the circular region of radius $r=r_0-a$.
Previous studies on the static packing of long-range repulsive
particles confined in the disk reveal the negative value for $Q$, indicating the
emergent hyperbolic geometry in the inhomogeneity created by the
long-range interaction~\cite{Mughal2007,yao2013topological,soni2018emergent}.
Here, we generalize the disk model from the static to the dynamical regime, and
explore the defect structure in the dynamical long-range interacting system.

We analyze hundreds of statistically independent equilibrium particle
configurations at varying $\lambda$, and obtain the histograms of the total
topological charge $Q$ for the cases of $\lambda/a=0.05$ and $\lambda/a=10$ as
shown in Figs.~\ref{fig_defect}(c) and \ref{fig_defect}(d). From
Fig.~\ref{fig_defect}(d) for $\lambda/a=10$, we see that in most equilibrium
states the total topological charge is negative. However, equilibrium states
with positive $Q$ also exist. In contrast, the value for $Q$ is negative in
the static ground state of long-range interacting disk
system~\cite{Mughal2007,yao2013topological,soni2018emergent}. As such, the
dynamical disk model supports richer defect structures.
Figure~\ref{fig_defect}(d) shows that the mean value for $Q$ is negative for
$\lambda/a=10$. In contrast, for the case of $\lambda/a=0.05$ in
Fig.~\ref{fig_defect}(c), the total topological charge is uniformly positive in
all of the equilibrium states. The dependence of $\langle Q \rangle$ on $\lambda$ is plotted in
Fig.~\ref{fig_defect}(b); we obtain the values for $\langle Q \rangle$ and the
error bars by analyzing over 200 statistically independent particle configurations in
equilibrium. Figure~\ref{fig_defect}(b) shows that the value for $\langle Q
\rangle$ turns negative with the increase of $\lambda$. Therefore, it is in the
sense of statistical averaging that the dynamical long-range interacting system
preserves the negative $\langle Q \rangle$-value and the hyperbolic geometry as in the static
system.

Finally, we briefly discuss the issue of boundary effect. The confining geometry
of the disk brings in the length scale $r_0$ (the radius of the disk). The
degree of the boundary effect may be measured by the ratio $\lambda/r_0$.
Calculations show that the value of $\lambda/r_0$ is as small as $0.02$ for
$\lambda/a = 0.75$, when $\langle Q \rangle$ becomes negative as shown in
Fig.~\ref{fig_defect}(b). As such, the phenomenon of negative $\langle Q \rangle$-value reflects
the intrinsic dynamical and statistical property of the long-range interacting
system. We also notice that the value of $\lambda/r_0$ increases from $0.02$
($\lambda/a = 0.75$) to $0.27$ ($\lambda/a =10$). To reduce the boundary effect
and meanwhile to retain the effect of long-range interaction, one shall work in
the regime of small $\lambda/r_0$ and large $\lambda/a$. This condition can be
fulfilled by specifying a large value for $N$, since
$\lambda/r_0=(\lambda/a)\times (\sqrt{2\pi/\sqrt{3}N})$. By striking a balance
of simulation time and system size, the maximum value for $N$ is $5000$ in this
work. Since the computation time is a quadratic function of $N$ for simulating
long-range interacting systems, it is desirable to employ efficient algorithm to
approach the large-$N$ limit for exploring the interested regime of small
$\lambda/r_0$ and large $\lambda/a$. This is beyond the scope of this work.

\section{Conclusion}

In summary, based on the screened Coulomb interacting disk model, we investigate
the dynamical and statistical effects of the long-range interaction by analyzing
the classical microscopic dynamics. Specifically, we reveal the fundamental
difference of the short- and long-range interactions in the aspects of energy
transfer mode, single-particle and collective dynamics, and the underlying
defect structure.  This work demonstrates the emergent dynamical and statistical
regularities in long-range interacting many-particle systems, and suggests
the rich physics arising from the interplay of long-range interaction,
topology and dynamics.

\begin{acknowledgements}
This work was supported by the National Natural Science Foundation of China
(Grants No. BC4190050).  The author thanks the support from the Student
Innovation Center at Shanghai Jiao Tong University.
\end{acknowledgements}

\begin{figure}[th]  
\centering 
\includegraphics[width=3.45in]{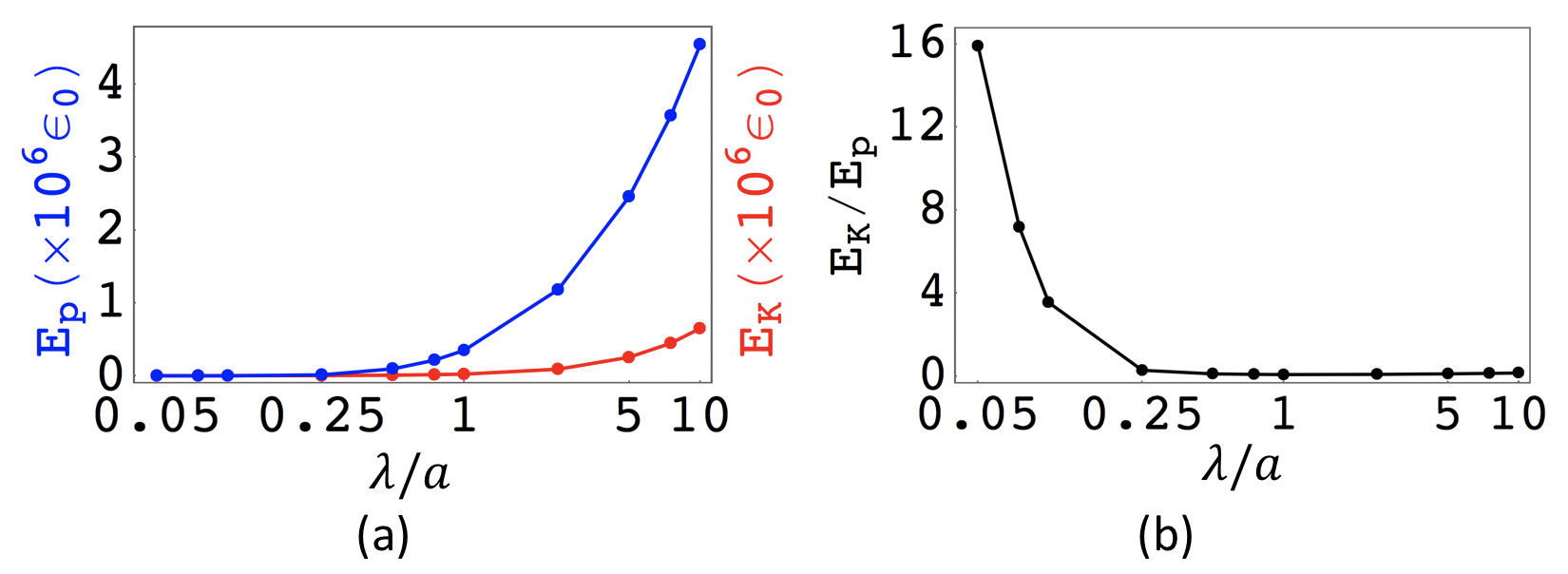}
  \caption{Plot of the potential energy $E_p$ and the kinetic energy $E_k$ versus
  $\lambda$ in equilibrium state. $N=5000$. $k_0=10^5$. $V_0=1$. }
\label{fig_energySI}
\end{figure}

\begin{figure*}[hb]  
\centering 
\includegraphics[width=7in]{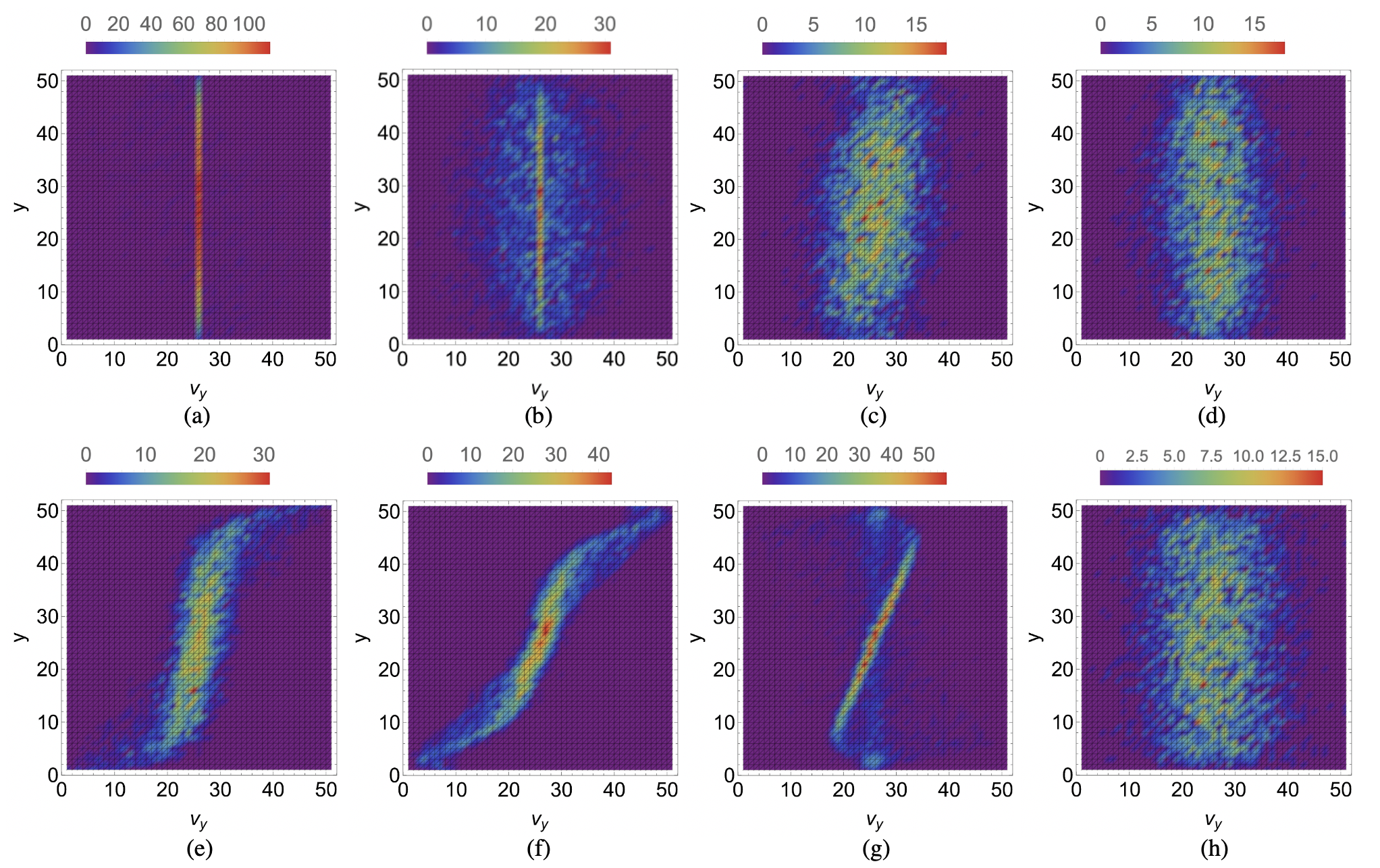}
\caption{Short- and long-range interacting particle systems exhibit distinct
  scenarios of dynamical evolution as represented in the space of $\{y,
  v_y\}$. (a)-(d) $\lambda/a=0.05$. $t/\tau_0=\{0.15, 0.90, 1.35, 5.85\}$.
  (e)-(h) $\lambda/a=10$. $t/\tau_0=\{0.0015, 0.009, 0.036, 6.0\}$. The space is divided into $50\times 50$ cells. The colored legends indicate the
  number of particles in $\delta y \delta v_y$. $y \in [-y_m, y_m]$ and $v_y \in
  [-v_m, v_m]$.  (a)-(d) $y_m= \{1.0, 0.9, 1.0, 1.0\}$. $v_m=\{1.92, 2.6, 2.83,
  3.24 \}$.  (e)-(h) $y_m= \{1.01, 1.12, 1.12, 1.15\}$. $v_m= \{16.40, 28.50,
  42.21, 59.23\}$. $k_0=10^5$. $N=5000$. $V_0=1$. }
\label{fig_yvy}
\end{figure*}

\newpage

\begin{figure*}[th]  
\centering 
\includegraphics[width=6in]{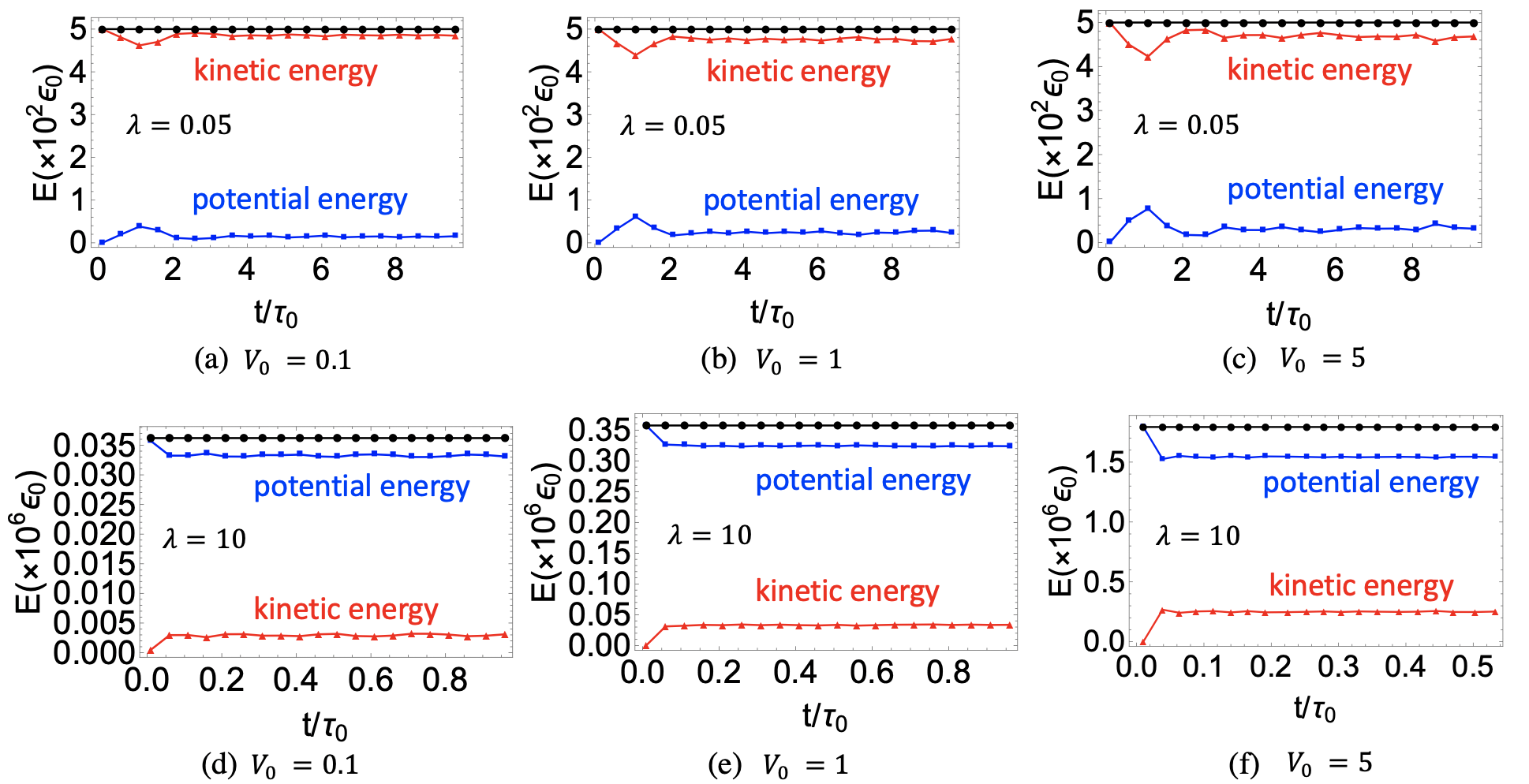}
  \caption{Temporal variation of the kinetic (red curves) and potential (blue
  curves) energies for typical short- and long-range interacting systems under
  varying strength $V_0$ of the screened Coulomb potential. $N=1000$.
  $k_0=10^5$. 
     }
\label{fig_energy_V0}
\end{figure*}

\begin{figure}[t]  
\centering 
\includegraphics[width=3.5in]{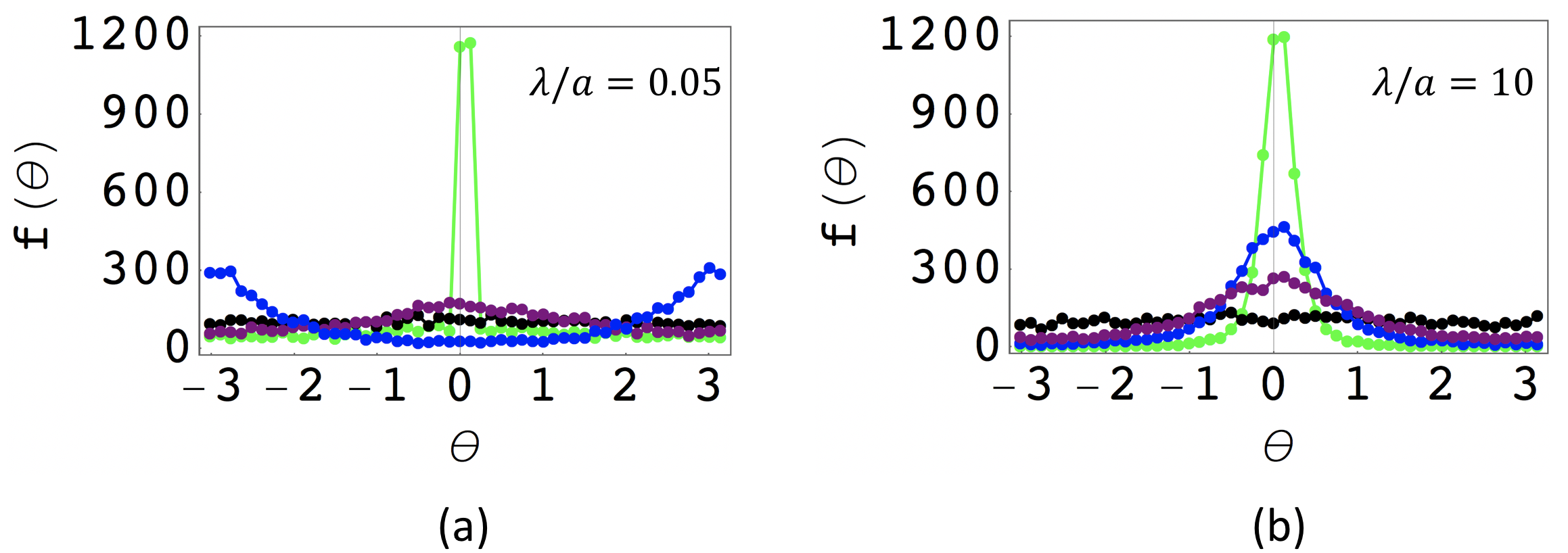}
  \caption{Relaxation of the orientation of the particle velocity for
  $\lambda/a=0.05$ (a) and $\lambda/a=10$ (b). $\theta$ is the angle between the
  direction of the particle velocity and x-axis. In (a), $t/\tau_0= 0.57$
  (green), 2.07 (blue), 3.57 (purple), and 5.85 (black). In (b), $t/\tau_0=
  0.0001$ (green), 0.0003 (blue), 0.0006 (purple), and 0.01 (black).  $N=5000$.
  $k_0=10^5$. $V_0=1$.  }
\label{fig_theta}
\end{figure}

\begin{figure}[h]  
\centering 
\includegraphics[width=3.5in]{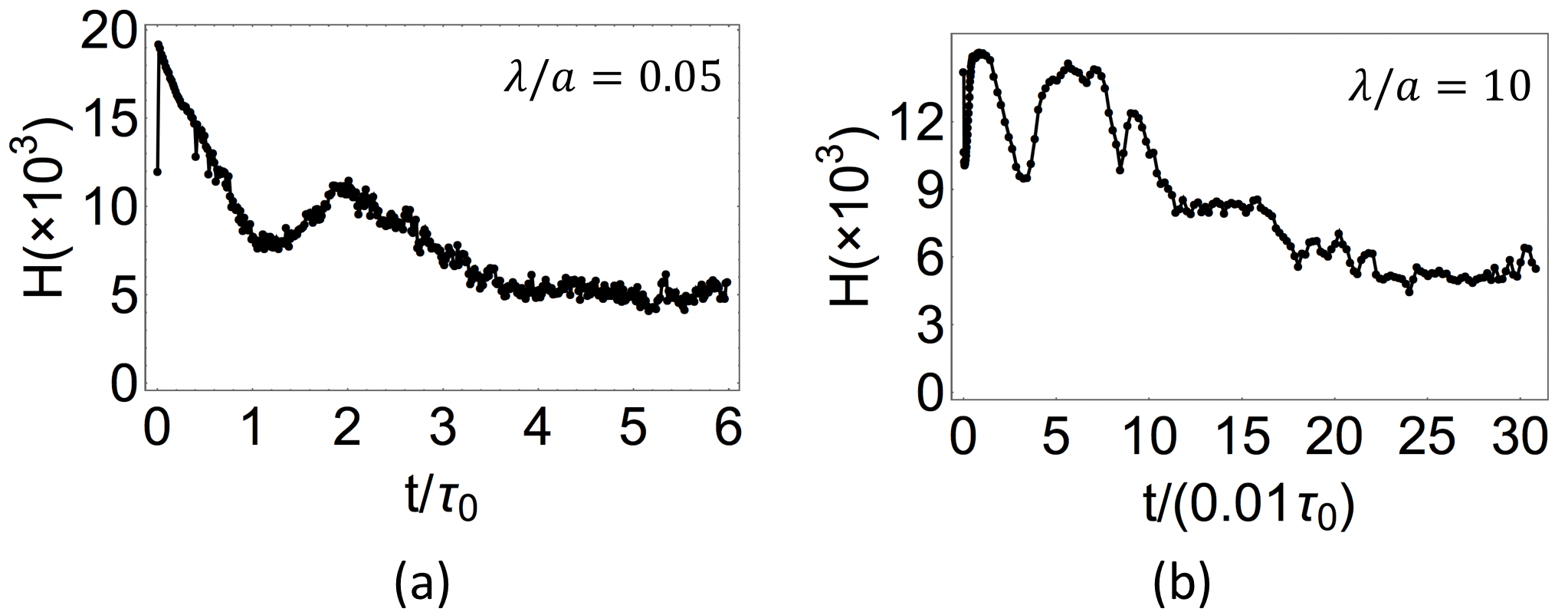}
  \caption{Temporal variation of the $H$ curve under a softer confining
  potential in comparison with that in the main text. $k_0=10^4$. $\lambda/a=0.05$
  (a) and $\lambda/a=10$ (b). The $H$ curve for $\lambda/a=10$ (b) takes longer time
  to become stable than that at $k_0=10^5$. In contrast, the $H$ curve for
  $\lambda/a=0.05$ (a) is almost identical to Fig.3a in the main text for
  $\lambda/a=0.05$ and $k_0=10^5$. $N=5000$. $V_0=1$. }
\label{fig_H}
\end{figure}

\begin{figure}[t]  
\centering 
\includegraphics[width=3.45in]{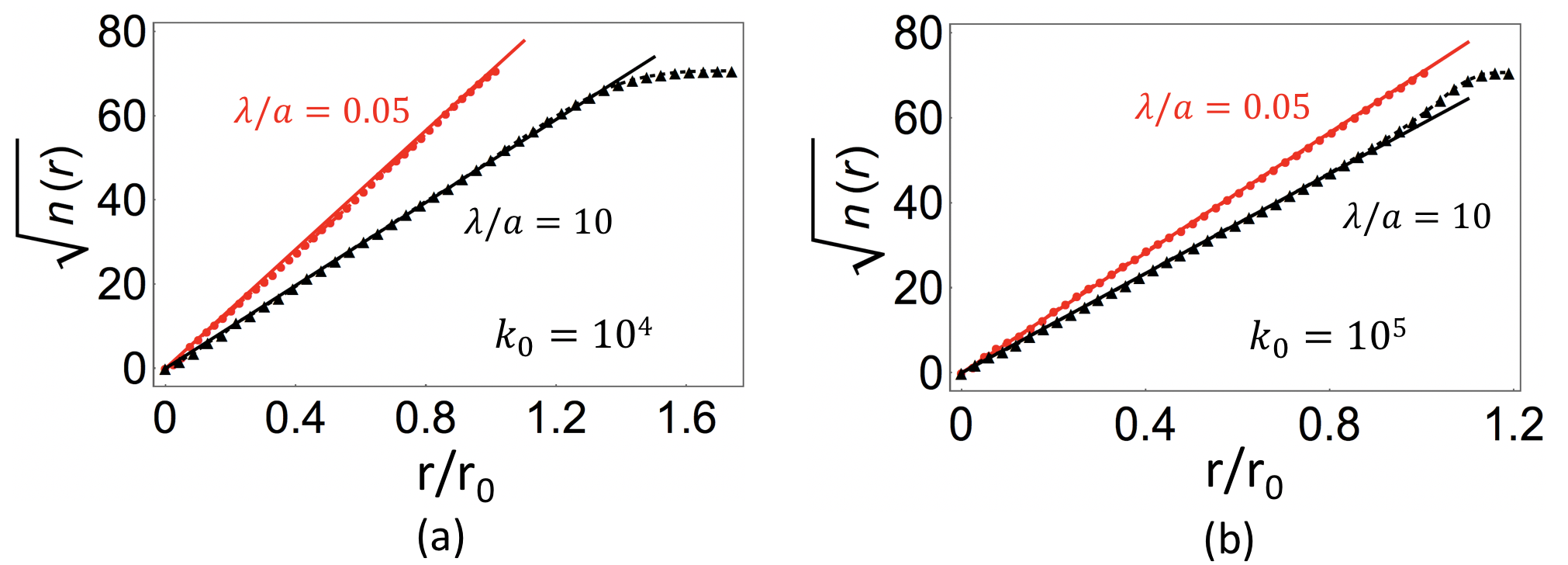}
  \caption{Plot of the square root of the cumulative particle distribution
  $\sqrt{n(r)}$ in equilibrium configurations. $k_0=10^4$ (a). $k_0=10^5$
  (b). The solid red lines represent the uniform distribution of particles.
  $N=5000$. $V_0=1$. }
\label{fig_density}
\end{figure}

\section*{Appendix A: Simulation details}

We employ the adaptive Verlet method to construct long-time particle
trajectories~\cite{rapaport2004art}. The time step $dt$ is dynamically varying
for striking a balance of the
energy conservation and computational efficiency. 

We denote the trajectory of any particle labelled $i$ as $\{ \vec{x}_i(t_j) \}$. From the initial state
specified by $\vec{x}_i(t_0)$ and $\dot{\vec{x}}_i(t_0)$, we obtain
$\vec{x}_i(t_i=t_0+h)$ by
\begin{eqnarray}
\vec{x}_i(t+h) = \vec{x}_i(t) + \dot{\vec{x}}_i(t)h  + \frac{1}{2}
\ddot{\vec{x}}_i(t) h^2 
    +{\cal{O}} (h^3).\nonumber
\end{eqnarray}
$\ddot{\vec{x}}_i(t)= \vec{F}_i(t)/m_0$, $\vec{F}_i(t)$ is the force on the
particle $i$ at time $t$, and $m_0$ is the mass of the particle. $\vec{F}_i=\sum_{i
\neq j} \vec{F}_{ij} + \vec{F}_{iW}$, where the first term is the interaction force from all
the other particles, and the second term arises if $|\vec{x}_i| > r_0$.
According to the adaptive Verlet method, from $\vec{x}_i(t_j)$ and
$\vec{x}_i(t_{j-1})$, we have
\begin{eqnarray}
  \vec{x}_i(t_{j+1}) = \vec{x}_i(t_j) + \left(\vec{x}_i(t_j) -
  \vec{x}_i(t_{j-1})\right)\frac{dt_{j}}{dt_{j-1}} 
   \nonumber \\
   + \ddot{\vec{x}}_i(t_j) \frac{dt_j+dt_{j-1}}{2} dt_j.
\end{eqnarray}
For uniform time step $dt_j=h$, the above equation reduces to the ordinary Verlet
integration scheme:
\begin{eqnarray}
\vec{x}_i(t+2h) = 2 \vec{x}_i(t+h) - \vec{x}_i(t) +
\ddot{\vec{x}}_i(t+h)h^2   +
{\cal{O}} (h^4).\nonumber
\end{eqnarray}

We employ the procedure of random disk packing to generate the initial random
configuration~\cite{lubachevsky1990geometric}.  Specifically, the disks of
radius $r_d$ are placed within the circle of radius $r_0$ in sequence. In this
process, each newly added disk shall not overlap any existent disk. The centers
of the disks constitute the initial positions of the point particles. Typically,
the value of $r_d$ is about $0.3 a$, where $a$ is the mean distance of nearest
particles. The reason of using random disk packing instead of random point
packing is as follows. Simulations show that random point packing could lead to
aggregation of particles. The resulting large force requires a very fine time
step to fulfill the conservation law of energy, which significantly slows
down the dynamical evolution of the system in simulations.

The total mechanical energy in our simulations is well conserved at a high
precision up to several decimal digits in the energy value. In the main text, we
have shown the temporal variation of the kinetic and potential energies. Here,
in Fig.~\ref{fig_energySI}, we plot the kinetic and potential energies versus
$\lambda$ in equilibrium state. From Fig.~\ref{fig_energySI}(a), we see that the
potential energy increases much faster than the kinetic energy. Furthermore,
Fig.~\ref{fig_energySI}(b) shows the rapid decline of the ratio $E_k/E_p$ with
the increase of $\lambda$.

\section*{Appendix B: More information about relaxation kinetics}

Typical instantaneous states in the dynamical evolution of
the system in the space spanned by $x$ and $v_x$ are
presented in Fig.~\ref{fig_phase} in the main text. 
In Fig.~\ref{fig_yvy}, we further present the identical dynamical evolutions as in
Fig.~\ref{fig_phase} in the complementary space of $\{y, v_y\}$. It is
observed that for the case of $\lambda/a=10$, the
patterns in both spaces of $\{x, v_x\}$ and $\{y, v_y\}$ become similar
after $t/\tau_0 = 0.009$ [see Fig.~\ref{fig_phase}(f) and
Fig.~\ref{fig_yvy}(f)]. In other words, the distribution function for
the system of $\lambda/a=10$ becomes symmetric with respect to $(x, v_x)$ and $(y, v_y)$ in a
much faster fashion in comparison with the system of $\lambda/a=0.05$.

In Fig.~\ref{fig_energy_V0}, we show the temporal variation of the kinetic and
potential energies for typical short- and long-range interacting systems under
varying $V_0$. The value of $V_0$ reflects the strength of the screened Coulomb
potential, as shown in Eq.(\ref{debye}). Since $V_0$ is measured in the unit of
$m_0 r_0 v_0^2$, varying $V_0$ is equivalent to changing the value of the
initial speed $v_0$. Figure~\ref{fig_energy_V0} shows that the conversion of
kinetic and potential energies in either short- or long-range interacting
systems conforms to a common scenario as the value of $V_0$ is varied from
$V_0=0.1$ to $V_0 = 5$. We also notice that unlike the case of $\lambda/a=10$
[the lower panels in Figs.~\ref{fig_energy_V0}(d)-\ref{fig_energy_V0}(f)], the
total energy of the systems with $\lambda/a=0.05$ is almost invariant as the
value of $V_0$ is varied. This could be attributed to the short-range nature
of the interaction, which resembles a hard repulsion at short distance.

In Figs.~\ref{fig_theta}(a) and \ref{fig_theta}(b), we present typical
instantaneous distributions of the orientation of the particle velocity in the
relaxation process for $\lambda/a=0.05$ and $\lambda/a=10$, respectively. $\theta$
is the angle between the direction of the particle velocity and x-axis. Similar
to the relaxation of particle speed, the relaxation of $\theta$ in the
long-range interacting system is also much faster than that in the short-range
interacting system.

In Figs.~\ref{fig_H}(a) and \ref{fig_H}(b), we show the temporal variation of the $H$
function under a softer confining potential for
$\lambda/a=0.05$ and $\lambda/a=10$, respectively. Here, $k_0=10^4$, which is ten
times less than the case we have discussed in the main text. In comparison
with the $H$-curves in Figs.3(a) and 3(b) in the main text, we find that a softer
confining potential tends to significantly slow the relaxation kinetics for the long-range
interacting system. The relaxation time increases from about $0.07\tau_0$ at $k_0=10^5$
to about $0.2\tau_0$ at $k_0=10^4$. In contrast, the relaxation rate of the short-range
interacting system is unaffected by the stiffness of the boundary wall.

\section*{Appendix C: Distribution of particle density in equilibrium}

Figures~\ref{fig_density}(a) and \ref{fig_density}(b) show the square root of
the cumulative particle distribution in equilibrium particle configurations.
$n(r)$ is the total number of particles inside the circle of radius $r$. We find
that the $\sqrt{n(r)}$ curve for $\lambda/a=0.05$ is linear in the interval $r/r_0
\in[0, 1]$ in both Figs.~\ref{fig_density}(a) and \ref{fig_density}(b), where
$r_0$ is the radius of the disk. The linearity of $\sqrt{n(r)}$ with $r$
indicated a uniform distribution of particles, since $\sqrt{n(r)} = \sqrt{\pi
\rho_0}r \propto r$ for a uniform particle distribution of density $\rho_0$.
Figure~\ref{fig_density} shows that changing the stiffness of the confining
potential leads to the variation of the particle density for the long-range
interacting system. More particles are accumulated near a softer boundary, which
reduces the particle density within the disk. In contrast, for the short-range
interacting system, the particle density distribution is almost unaffected by
the stiffness of the confining potential.


\begin{thebibliography}{39}
\expandafter\ifx\csname natexlab\endcsname\relax\def\natexlab#1{#1}\fi
\expandafter\ifx\csname bibnamefont\endcsname\relax
  \def\bibnamefont#1{#1}\fi
\expandafter\ifx\csname bibfnamefont\endcsname\relax
  \def\bibfnamefont#1{#1}\fi
\expandafter\ifx\csname citenamefont\endcsname\relax
  \def\citenamefont#1{#1}\fi
\expandafter\ifx\csname url\endcsname\relax
  \def\url#1{\texttt{#1}}\fi
\expandafter\ifx\csname urlprefix\endcsname\relax\def\urlprefix{URL }\fi
\providecommand{\bibinfo}[2]{#2}
\providecommand{\eprint}[2][]{\url{#2}}

\bibitem[{\citenamefont{Cercignani et~al.}(1998)}]{cercignani1998ludwig}
\bibinfo{author}{\bibfnamefont{C.}~\bibnamefont{Cercignani}}
  \bibnamefont{et~al.}, \emph{\bibinfo{title}{Ludwig Boltzmann: The Man Who
  Trusted Atoms}} (\bibinfo{publisher}{Oxford University Press, Oxford},
  \bibinfo{year}{1998}).

\bibitem[{\citenamefont{Boltzmann}(1964)}]{boltzmann1964lectures}
\bibinfo{author}{\bibfnamefont{L.}~\bibnamefont{Boltzmann}},
  \emph{\bibinfo{title}{Lectures On Gas Theory}}
  (\bibinfo{publisher}{University of California Press, Berkeley},
  \bibinfo{year}{1964}).

\bibitem[{\citenamefont{Ma}(1985)}]{Ma1985}
\bibinfo{author}{\bibfnamefont{S.}~\bibnamefont{Ma}},
  \emph{\bibinfo{title}{Statistical Mechanics}} (\bibinfo{publisher}{World
  Scientific, Singapore}, \bibinfo{year}{1985}).

\bibitem[{\citenamefont{Ehrenfest and
  Ehrenfest}(2002)}]{ehrenfest2002conceptual}
\bibinfo{author}{\bibfnamefont{P.}~\bibnamefont{Ehrenfest}} \bibnamefont{and}
  \bibinfo{author}{\bibfnamefont{T.}~\bibnamefont{Ehrenfest}},
  \emph{\bibinfo{title}{The Conceptual Foundations of The Statistical Approach
  in Mechanics}} (\bibinfo{publisher}{Courier Corporation, Massachusetts},
  \bibinfo{year}{2002}).

\bibitem[{\citenamefont{Frenkel and Lou{\"e}t}(2016)}]{frenkel2016interview}
\bibinfo{author}{\bibfnamefont{D.}~\bibnamefont{Frenkel}} \bibnamefont{and}
  \bibinfo{author}{\bibfnamefont{S.}~\bibnamefont{Lou{\"e}t}},
  \bibinfo{journal}{Eur. Phys. J. E} \textbf{\bibinfo{volume}{39}},
  \bibinfo{pages}{68} (\bibinfo{year}{2016}).

\bibitem[{\citenamefont{Campa et~al.}(2014)\citenamefont{Campa, Dauxois,
  Fanelli, and Ruffo}}]{campa2014physics}
\bibinfo{author}{\bibfnamefont{A.}~\bibnamefont{Campa}},
  \bibinfo{author}{\bibfnamefont{T.}~\bibnamefont{Dauxois}},
  \bibinfo{author}{\bibfnamefont{D.}~\bibnamefont{Fanelli}}, \bibnamefont{and}
  \bibinfo{author}{\bibfnamefont{S.}~\bibnamefont{Ruffo}},
  \emph{\bibinfo{title}{Physics of Long-Range Interacting Systems}}
  (\bibinfo{publisher}{Oxford University Press, Oxford, UK},
  \bibinfo{year}{2014}).

\bibitem[{\citenamefont{Levin et~al.}(2014)\citenamefont{Levin, Pakter,
  Rizzato, Teles, and Benetti}}]{levin2014nonequilibrium}
\bibinfo{author}{\bibfnamefont{Y.}~\bibnamefont{Levin}},
  \bibinfo{author}{\bibfnamefont{R.}~\bibnamefont{Pakter}},
  \bibinfo{author}{\bibfnamefont{F.~B.} \bibnamefont{Rizzato}},
  \bibinfo{author}{\bibfnamefont{T.~N.} \bibnamefont{Teles}}, \bibnamefont{and}
  \bibinfo{author}{\bibfnamefont{F.~P.} \bibnamefont{Benetti}},
  \bibinfo{journal}{Phys. Rep.} \textbf{\bibinfo{volume}{535}},
  \bibinfo{pages}{1} (\bibinfo{year}{2014}).

\bibitem[{\citenamefont{Pakter and Levin}(2017)}]{pakter2017entropy}
\bibinfo{author}{\bibfnamefont{R.}~\bibnamefont{Pakter}} \bibnamefont{and}
  \bibinfo{author}{\bibfnamefont{Y.}~\bibnamefont{Levin}}, \bibinfo{journal}{J.
  Stat. Mech: Theory Exp.} \textbf{\bibinfo{volume}{2017}},
  \bibinfo{pages}{044001} (\bibinfo{year}{2017}).

\bibitem[{\citenamefont{Lighthill}(1976)}]{lighthill1976flagellar}
\bibinfo{author}{\bibfnamefont{J.}~\bibnamefont{Lighthill}},
  \bibinfo{journal}{SIAM Rev} \textbf{\bibinfo{volume}{18}},
  \bibinfo{pages}{161} (\bibinfo{year}{1976}).

\bibitem[{\citenamefont{Chattopadhyay and Wu}(2009)}]{chattopadhyay2009effect}
\bibinfo{author}{\bibfnamefont{S.}~\bibnamefont{Chattopadhyay}}
  \bibnamefont{and} \bibinfo{author}{\bibfnamefont{X.-L.} \bibnamefont{Wu}},
  \bibinfo{journal}{Biophys. J.} \textbf{\bibinfo{volume}{96}},
  \bibinfo{pages}{2023} (\bibinfo{year}{2009}).

\bibitem[{\citenamefont{Tabi et~al.}(2010)\citenamefont{Tabi, Mohamadou, and
  Kofan{\'e}}}]{tabi2010long}
\bibinfo{author}{\bibfnamefont{C.}~\bibnamefont{Tabi}},
  \bibinfo{author}{\bibfnamefont{A.}~\bibnamefont{Mohamadou}},
  \bibnamefont{and}
  \bibinfo{author}{\bibfnamefont{T.}~\bibnamefont{Kofan{\'e}}},
  \bibinfo{journal}{Eur. Phys. J. E} \textbf{\bibinfo{volume}{32}},
  \bibinfo{pages}{327} (\bibinfo{year}{2010}).

\bibitem[{\citenamefont{Dallaston et~al.}(2018)\citenamefont{Dallaston,
  Fontelos, Tseluiko, and Kalliadasis}}]{dallaston2018discrete}
\bibinfo{author}{\bibfnamefont{M.~C.} \bibnamefont{Dallaston}},
  \bibinfo{author}{\bibfnamefont{M.~A.} \bibnamefont{Fontelos}},
  \bibinfo{author}{\bibfnamefont{D.}~\bibnamefont{Tseluiko}}, \bibnamefont{and}
  \bibinfo{author}{\bibfnamefont{S.}~\bibnamefont{Kalliadasis}},
  \bibinfo{journal}{Phys. Rev. Lett.} \textbf{\bibinfo{volume}{120}},
  \bibinfo{pages}{034505} (\bibinfo{year}{2018}).

\bibitem[{\citenamefont{Yao}(2019)}]{yao2019command}
\bibinfo{author}{\bibfnamefont{Z.}~\bibnamefont{Yao}}, \bibinfo{journal}{Phys.
  Rev. Lett.} \textbf{\bibinfo{volume}{122}}, \bibinfo{pages}{228002}
  (\bibinfo{year}{2019}).

\bibitem[{\citenamefont{Holm et~al.}(2001)\citenamefont{Holm, K{\'e}kicheff,
  and Podgornik}}]{Holm2001}
\bibinfo{author}{\bibfnamefont{C.}~\bibnamefont{Holm}},
  \bibinfo{author}{\bibfnamefont{P.}~\bibnamefont{K{\'e}kicheff}},
  \bibnamefont{and}
  \bibinfo{author}{\bibfnamefont{R.}~\bibnamefont{Podgornik}},
  \emph{\bibinfo{title}{Electrostatic Effects in Soft Matter and Biophysics}}
  (\bibinfo{publisher}{Springer, Berlin}, \bibinfo{year}{2001}).

\bibitem[{\citenamefont{Levin}(2002)}]{Levin2002}
\bibinfo{author}{\bibfnamefont{Y.}~\bibnamefont{Levin}}, \bibinfo{journal}{Rep.
  Prog. Phys.} \textbf{\bibinfo{volume}{65}}, \bibinfo{pages}{1577}
  (\bibinfo{year}{2002}).

\bibitem[{\citenamefont{Walker et~al.}(2011)\citenamefont{Walker, Kowalczyk,
  Olvera de~la Cruz, and Grzybowski}}]{Walker2011}
\bibinfo{author}{\bibfnamefont{D.~A.} \bibnamefont{Walker}},
  \bibinfo{author}{\bibfnamefont{B.}~\bibnamefont{Kowalczyk}},
  \bibinfo{author}{\bibfnamefont{M.}~\bibnamefont{Olvera de~la Cruz}},
  \bibnamefont{and} \bibinfo{author}{\bibfnamefont{B.~A.}
  \bibnamefont{Grzybowski}}, \bibinfo{journal}{Nanoscale}
  \textbf{\bibinfo{volume}{3}}, \bibinfo{pages}{1316} (\bibinfo{year}{2011}).

\bibitem[{\citenamefont{Toor et~al.}(2016)\citenamefont{Toor, Feng, and
  Russell}}]{toor2016self}
\bibinfo{author}{\bibfnamefont{A.}~\bibnamefont{Toor}},
  \bibinfo{author}{\bibfnamefont{T.}~\bibnamefont{Feng}}, \bibnamefont{and}
  \bibinfo{author}{\bibfnamefont{T.~P.} \bibnamefont{Russell}},
  \bibinfo{journal}{Eur. Phys. J. E} \textbf{\bibinfo{volume}{39}},
  \bibinfo{pages}{1} (\bibinfo{year}{2016}).

\bibitem[{\citenamefont{Gao et~al.}(2019)\citenamefont{Gao, Kewalramani,
  Valencia, Li, McCourt, Olvera de~la Cruz, and Bedzyk}}]{gao2019electrostatic}
\bibinfo{author}{\bibfnamefont{C.}~\bibnamefont{Gao}},
  \bibinfo{author}{\bibfnamefont{S.}~\bibnamefont{Kewalramani}},
  \bibinfo{author}{\bibfnamefont{D.~M.} \bibnamefont{Valencia}},
  \bibinfo{author}{\bibfnamefont{H.}~\bibnamefont{Li}},
  \bibinfo{author}{\bibfnamefont{J.~M.} \bibnamefont{McCourt}},
  \bibinfo{author}{\bibfnamefont{M.}~\bibnamefont{Olvera de~la Cruz}},
  \bibnamefont{and} \bibinfo{author}{\bibfnamefont{M.~J.}
  \bibnamefont{Bedzyk}}, \bibinfo{journal}{Proc. Natl. Acad. Sci. U.S.A.}
  \textbf{\bibinfo{volume}{116}}, \bibinfo{pages}{22030}
  (\bibinfo{year}{2019}).

\bibitem[{\citenamefont{Rapaport}(2004)}]{rapaport2004art}
\bibinfo{author}{\bibfnamefont{D.}~\bibnamefont{Rapaport}},
  \emph{\bibinfo{title}{The Art of Molecular Dynamics Simulation}}
  (\bibinfo{publisher}{Cambridge University Press, Cambridge, UK},
  \bibinfo{year}{2004}).

\bibitem[{\citenamefont{Grzybowski et~al.}(2003)\citenamefont{Grzybowski,
  Winkleman, Wiles, Brumer, and Whitesides}}]{grzybowski2003electrostatic}
\bibinfo{author}{\bibfnamefont{B.~A.} \bibnamefont{Grzybowski}},
  \bibinfo{author}{\bibfnamefont{A.}~\bibnamefont{Winkleman}},
  \bibinfo{author}{\bibfnamefont{J.~A.} \bibnamefont{Wiles}},
  \bibinfo{author}{\bibfnamefont{Y.}~\bibnamefont{Brumer}}, \bibnamefont{and}
  \bibinfo{author}{\bibfnamefont{G.~M.} \bibnamefont{Whitesides}},
  \bibinfo{journal}{Nat. Mater.} \textbf{\bibinfo{volume}{2}},
  \bibinfo{pages}{241} (\bibinfo{year}{2003}).

\bibitem[{\citenamefont{Vernizzi et~al.}(2011)\citenamefont{Vernizzi,
  Guerrero-Garc{\'\i}a, and Olvera de~la Cruz}}]{vernizzi2011coulomb}
\bibinfo{author}{\bibfnamefont{G.}~\bibnamefont{Vernizzi}},
  \bibinfo{author}{\bibfnamefont{G.~I.} \bibnamefont{Guerrero-Garc{\'\i}a}},
  \bibnamefont{and} \bibinfo{author}{\bibfnamefont{M.}~\bibnamefont{Olvera
  de~la Cruz}}, \bibinfo{journal}{Phys. Rev. E} \textbf{\bibinfo{volume}{84}},
  \bibinfo{pages}{016707} (\bibinfo{year}{2011}).

\bibitem[{\citenamefont{Lindgren et~al.}(2018)\citenamefont{Lindgren, Derbenev,
  Khachatourian, Chan, Stace, and Besley}}]{lindgren2018electrostatic}
\bibinfo{author}{\bibfnamefont{E.~B.} \bibnamefont{Lindgren}},
  \bibinfo{author}{\bibfnamefont{I.~N.} \bibnamefont{Derbenev}},
  \bibinfo{author}{\bibfnamefont{A.}~\bibnamefont{Khachatourian}},
  \bibinfo{author}{\bibfnamefont{H.-K.} \bibnamefont{Chan}},
  \bibinfo{author}{\bibfnamefont{A.~J.} \bibnamefont{Stace}}, \bibnamefont{and}
  \bibinfo{author}{\bibfnamefont{E.}~\bibnamefont{Besley}},
  \bibinfo{journal}{J. Chem. Theory Comput.} \textbf{\bibinfo{volume}{14}},
  \bibinfo{pages}{905} (\bibinfo{year}{2018}).

\bibitem[{\citenamefont{Mughal and Moore}(2007)}]{Mughal2007}
\bibinfo{author}{\bibfnamefont{A.}~\bibnamefont{Mughal}} \bibnamefont{and}
  \bibinfo{author}{\bibfnamefont{M.}~\bibnamefont{Moore}},
  \bibinfo{journal}{Phys. Rev. E} \textbf{\bibinfo{volume}{76}},
  \bibinfo{pages}{011606} (\bibinfo{year}{2007}).

\bibitem[{\citenamefont{Yao and Olvera de~la Cruz}(2013)}]{yao2013topological}
\bibinfo{author}{\bibfnamefont{Z.}~\bibnamefont{Yao}} \bibnamefont{and}
  \bibinfo{author}{\bibfnamefont{M.}~\bibnamefont{Olvera de~la Cruz}},
  \bibinfo{journal}{Phys. Rev. Lett.} \textbf{\bibinfo{volume}{111}},
  \bibinfo{pages}{115503} (\bibinfo{year}{2013}).

\bibitem[{\citenamefont{Soni et~al.}(2018)\citenamefont{Soni, G{\'o}mez, and
  Irvine}}]{soni2018emergent}
\bibinfo{author}{\bibfnamefont{V.}~\bibnamefont{Soni}},
  \bibinfo{author}{\bibfnamefont{L.~R.} \bibnamefont{G{\'o}mez}},
  \bibnamefont{and} \bibinfo{author}{\bibfnamefont{W.~T.}
  \bibnamefont{Irvine}}, \bibinfo{journal}{Phys. Rev. X}
  \textbf{\bibinfo{volume}{8}}, \bibinfo{pages}{011039} (\bibinfo{year}{2018}).

\bibitem[{\citenamefont{Debye}(1923)}]{debye1923theory}
\bibinfo{author}{\bibfnamefont{P.}~\bibnamefont{Debye}},
  \bibinfo{journal}{Physikalische Zeitschrift} \textbf{\bibinfo{volume}{24}},
  \bibinfo{pages}{185} (\bibinfo{year}{1923}).

\bibitem[{\citenamefont{Dobrynin and Rubinstein}(2005)}]{Dobrynin2005}
\bibinfo{author}{\bibfnamefont{A.}~\bibnamefont{Dobrynin}} \bibnamefont{and}
  \bibinfo{author}{\bibfnamefont{M.}~\bibnamefont{Rubinstein}},
  \bibinfo{journal}{Prog. Polym. Sci.} \textbf{\bibinfo{volume}{30}},
  \bibinfo{pages}{1049} (\bibinfo{year}{2005}).

\bibitem[{\citenamefont{Nelson}(2002)}]{nelson2002defects}
\bibinfo{author}{\bibfnamefont{D.~R.} \bibnamefont{Nelson}},
  \emph{\bibinfo{title}{Defects and Geometry in Condensed Matter Physics}}
  (\bibinfo{publisher}{Cambridge University Press, Cambridge},
  \bibinfo{year}{2002}).

\bibitem[{\citenamefont{Chen et~al.}(2020)\citenamefont{Chen, Yao, Tang, Tong,
  Yanagishima, Tanaka, and Tan}}]{chen2020morphology}
\bibinfo{author}{\bibfnamefont{Y.}~\bibnamefont{Chen}},
  \bibinfo{author}{\bibfnamefont{Z.}~\bibnamefont{Yao}},
  \bibinfo{author}{\bibfnamefont{S.}~\bibnamefont{Tang}},
  \bibinfo{author}{\bibfnamefont{H.}~\bibnamefont{Tong}},
  \bibinfo{author}{\bibfnamefont{T.}~\bibnamefont{Yanagishima}},
  \bibinfo{author}{\bibfnamefont{H.}~\bibnamefont{Tanaka}}, \bibnamefont{and}
  \bibinfo{author}{\bibfnamefont{P.}~\bibnamefont{Tan}}, \bibinfo{journal}{Nat.
  Phys.} \textbf{\bibinfo{volume}{17}}, \bibinfo{pages}{121}
  (\bibinfo{year}{2020}).

\bibitem[{\citenamefont{Lubachevsky and
  Stillinger}(1990)}]{lubachevsky1990geometric}
\bibinfo{author}{\bibfnamefont{B.~D.} \bibnamefont{Lubachevsky}}
  \bibnamefont{and} \bibinfo{author}{\bibfnamefont{F.~H.}
  \bibnamefont{Stillinger}}, \bibinfo{journal}{Journal of statistical Physics}
  \textbf{\bibinfo{volume}{60}}, \bibinfo{pages}{561} (\bibinfo{year}{1990}).

\bibitem[{\citenamefont{Kosterlitz and Thouless}(1973)}]{Kosterlitz1973}
\bibinfo{author}{\bibfnamefont{J.~M.} \bibnamefont{Kosterlitz}}
  \bibnamefont{and} \bibinfo{author}{\bibfnamefont{D.~J.}
  \bibnamefont{Thouless}}, \bibinfo{journal}{J. Phys. C: Solid State Phys.}
  \textbf{\bibinfo{volume}{6}}, \bibinfo{pages}{1181} (\bibinfo{year}{1973}).

\bibitem[{\citenamefont{Halperin and Nelson}(1978)}]{halperin1978theory}
\bibinfo{author}{\bibfnamefont{B.}~\bibnamefont{Halperin}} \bibnamefont{and}
  \bibinfo{author}{\bibfnamefont{D.~R.} \bibnamefont{Nelson}},
  \bibinfo{journal}{Phys. Rev. Lett.} \textbf{\bibinfo{volume}{41}},
  \bibinfo{pages}{121} (\bibinfo{year}{1978}).

\bibitem[{\citenamefont{Irvine et~al.}(2012)\citenamefont{Irvine, Bowick, and
  Chaikin}}]{irvine2012fractionalization}
\bibinfo{author}{\bibfnamefont{W.~T.} \bibnamefont{Irvine}},
  \bibinfo{author}{\bibfnamefont{M.~J.} \bibnamefont{Bowick}},
  \bibnamefont{and} \bibinfo{author}{\bibfnamefont{P.~M.}
  \bibnamefont{Chaikin}}, \bibinfo{journal}{Nat. Mater.}
  \textbf{\bibinfo{volume}{11}}, \bibinfo{pages}{948} (\bibinfo{year}{2012}).

\bibitem[{\citenamefont{Yao and Olvera de~la
  Cruz}(2014)}]{yao2014polydispersity}
\bibinfo{author}{\bibfnamefont{Z.}~\bibnamefont{Yao}} \bibnamefont{and}
  \bibinfo{author}{\bibfnamefont{M.}~\bibnamefont{Olvera de~la Cruz}},
  \bibinfo{journal}{Proc. Natl. Acad. Sci. U.S.A.}
  \textbf{\bibinfo{volume}{111}}, \bibinfo{pages}{5094} (\bibinfo{year}{2014}).

\bibitem[{\citenamefont{Kamien and Nelson}(1995)}]{kamien1995iterated}
\bibinfo{author}{\bibfnamefont{R.~D.} \bibnamefont{Kamien}} \bibnamefont{and}
  \bibinfo{author}{\bibfnamefont{D.~R.} \bibnamefont{Nelson}},
  \bibinfo{journal}{Phys. Rev. Lett.} \textbf{\bibinfo{volume}{74}},
  \bibinfo{pages}{2499} (\bibinfo{year}{1995}).

\bibitem[{\citenamefont{Grason}(2010)}]{Grason2010}
\bibinfo{author}{\bibfnamefont{G.~M.} \bibnamefont{Grason}},
  \bibinfo{journal}{Phys. Rev. Lett.} \textbf{\bibinfo{volume}{105}},
  \bibinfo{pages}{045502} (\bibinfo{year}{2010}).

\bibitem[{\citenamefont{Lidmar et~al.}(2003)\citenamefont{Lidmar, Mirny, and
  Nelson}}]{lidmar2003virus}
\bibinfo{author}{\bibfnamefont{J.}~\bibnamefont{Lidmar}},
  \bibinfo{author}{\bibfnamefont{L.}~\bibnamefont{Mirny}}, \bibnamefont{and}
  \bibinfo{author}{\bibfnamefont{D.~R.} \bibnamefont{Nelson}},
  \bibinfo{journal}{Phys. Rev. E} \textbf{\bibinfo{volume}{68}},
  \bibinfo{pages}{051910} (\bibinfo{year}{2003}).

\bibitem[{\citenamefont{Marchetti et~al.}(2013)\citenamefont{Marchetti, Joanny,
  Ramaswamy, Liverpool, Prost, Rao, and Simha}}]{marchetti2013hydrodynamics}
\bibinfo{author}{\bibfnamefont{M.}~\bibnamefont{Marchetti}},
  \bibinfo{author}{\bibfnamefont{J.}~\bibnamefont{Joanny}},
  \bibinfo{author}{\bibfnamefont{S.}~\bibnamefont{Ramaswamy}},
  \bibinfo{author}{\bibfnamefont{T.}~\bibnamefont{Liverpool}},
  \bibinfo{author}{\bibfnamefont{J.}~\bibnamefont{Prost}},
  \bibinfo{author}{\bibfnamefont{M.}~\bibnamefont{Rao}}, \bibnamefont{and}
  \bibinfo{author}{\bibfnamefont{R.~A.} \bibnamefont{Simha}},
  \bibinfo{journal}{Rev. Mod. Phys.} \textbf{\bibinfo{volume}{85}},
  \bibinfo{pages}{1143} (\bibinfo{year}{2013}).

\bibitem[{\citenamefont{Keber et~al.}(2014)\citenamefont{Keber, Loiseau,
  Sanchez, DeCamp, Giomi, Bowick, Marchetti, Dogic, and
  Bausch}}]{keber2014topology}
\bibinfo{author}{\bibfnamefont{F.~C.} \bibnamefont{Keber}},
  \bibinfo{author}{\bibfnamefont{E.}~\bibnamefont{Loiseau}},
  \bibinfo{author}{\bibfnamefont{T.}~\bibnamefont{Sanchez}},
  \bibinfo{author}{\bibfnamefont{S.~J.} \bibnamefont{DeCamp}},
  \bibinfo{author}{\bibfnamefont{L.}~\bibnamefont{Giomi}},
  \bibinfo{author}{\bibfnamefont{M.~J.} \bibnamefont{Bowick}},
  \bibinfo{author}{\bibfnamefont{M.~C.} \bibnamefont{Marchetti}},
  \bibinfo{author}{\bibfnamefont{Z.}~\bibnamefont{Dogic}}, \bibnamefont{and}
  \bibinfo{author}{\bibfnamefont{A.~R.} \bibnamefont{Bausch}},
  \bibinfo{journal}{Science} \textbf{\bibinfo{volume}{345}},
  \bibinfo{pages}{1135} (\bibinfo{year}{2014}).

\end{thebibliography}

\end{document}